\documentclass[conference]{IEEEtran}

\usepackage[utf8]{inputenc}
\usepackage{algorithmic}
\usepackage{amsmath,amssymb,amsfonts}
\usepackage{booktabs}
\usepackage{graphicx}
\usepackage{hyperref}
\usepackage[draft]{minted}
\usepackage{pgfplots}
\usepackage{pmboxdraw}
\usepackage{textcomp}
\usepackage{tikz}
\usepackage{xcolor}
\usepackage{subfig}
\usepackage{diagbox}

\begin{document}

\title{Better Safe Than Sorry! Automated Identification of Functionality-Breaking Security-Configuration Rules}

\author{
\IEEEauthorblockN{Patrick Stöckle\IEEEauthorrefmark{1}, Michael Sammereier\IEEEauthorrefmark{1}, Bernd Grobauer\IEEEauthorrefmark{2} and Alexander Pretschner\IEEEauthorrefmark{1}}
\IEEEauthorblockA{\IEEEauthorrefmark{1}Chair of Software and Systems Engineering \\
Technical University of Munich (TUM)\\
Munich, Germany \\
Email: \href{mailto:patrick.stoeckle@tum.de,michael.sammereier@tum.de,alexander.pretschner@tum.de}{\{patrick.stoeckle, michael.sammereier, alexander.pretschner\}@tum.de}\\
ORCID: \href{https://orcid.org/0000-0003-0193-5871}{0000-0003-0193-5871}, \href{https://orcid.org/0000-0002-5786-9554}{0000-0002-5786-9554}, \href{https://orcid.org/0000-0002-5573-1201}{0000-0002-5573-1201}
}
\IEEEauthorblockA{\IEEEauthorrefmark{2}T CST\\
Siemens AG\\
Munich, Germany \\
Email: \href{mailto:bernd.grobauer@siemens.de}{bernd.grobauer@siemens.de}, ORCID: \href{https://orcid.org/0000-0003-0792-3935}{0000-0003-0792-3935}}
}

\providecommand*{\listingautorefname}{Listing}
\providecommand*{\listingname}{Listing}
\providecommand*{\sectionautorefname}{Section}
\newcommand{\cisLong}{Center for Internet Security}
\newcommand{\cisShort}{CIS}
\newcommand{\tumLong}{Technical University of Munich}
\newcommand{\tumShort}{TUM}
\newcommand{\ctShort}{combinatoria}
\newcommand{\ctLong}{Combinatorial Testing}
\newcommand{\caShort}{CA}
\newcommand{\caShorts}{covering arrays}
\newcommand{\caLong}{Covering Array}
\newcommand{\actsShort}{ACTS}
\newcommand{\actsLong}{Automated Combinatorial Testing for Software}

\maketitle

\begin{abstract}
Insecure default values in software settings can be exploited by attackers to compromise the system that runs the software.
As a countermeasure, there exist security-configuration guides specifying in detail which values are secure.
However, most administrators still refrain from hardening existing systems because the system functionality is feared to deteriorate if secure settings are applied.
To foster the application of security-configuration guides, it is necessary to identify those rules that would restrict the functionality.

This article presents our approach to use combinatorial testing to find problematic combinations of rules and machine learning techniques to identify the problematic rules within these combinations.
The administrators can then apply only the unproblematic rules and, therefore, increase the system's security without the risk of disrupting its functionality.
To demonstrate the usefulness of our approach, we applied it to real-world problems drawn from discussions with administrators at Siemens and found the problematic rules in these cases.
We hope that this approach and its open-source implementation motivate more administrators to harden their systems and, thus, increase their systems' general security.
\end{abstract}


\section{Introduction}
\label{sec:TEST:introduction}

\begin{figure*}[t]
\centering
\includegraphics[width=\textwidth]{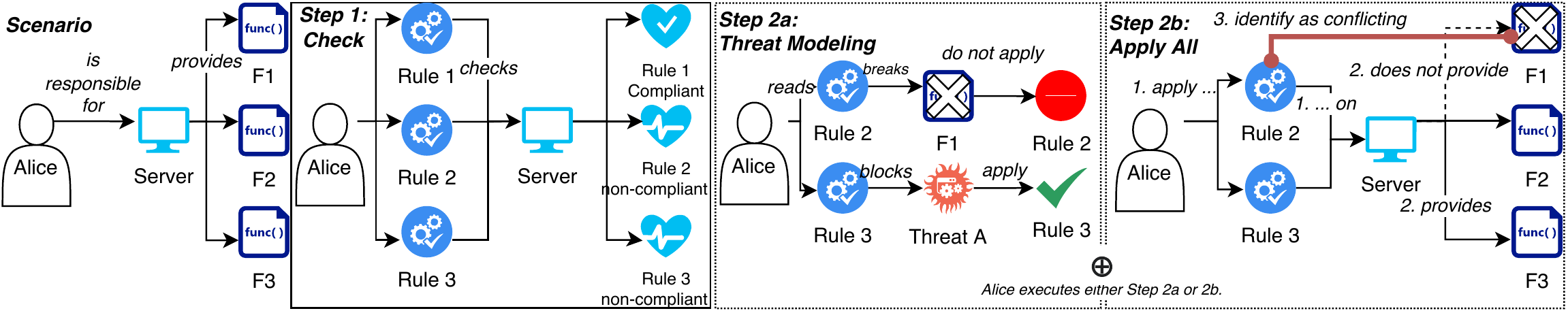}
\caption{The current process of hardening existing systems.}
\label{fig:TEST:state-of-the-art}
\end{figure*}

Researchers showed that in 15\% of all data breaches, the attackers exploited cloud misconfigurations~\cite{security2022cost}.
Nevertheless, there are various factors -- mainly lack of knowledge -- that prevent administrators from configuring their systems securely~\cite{dietrich2018investigating}.
For most systems, we can resolve the \textit{Lack of knowledge} by using a security-configuration guide from independent organizations like the \cisLong{} (\cisShort{}).
A guide is a set of rules and every rule specifies to which value the administrator has to set a configuration setting to make the system more secure.
However, administrators are very reluctant to apply this configuration hardening to systems in production.
Even if a guide is available and the necessary processes and tools to implement guides efficiently, they are discouraged by the fear of breaking the existing functionality.
In this article, we focus on this problem of hardening existing systems and detect functionality-breaking rules automatically.

\subsection{Motivating Example}

\autoref{fig:TEST:state-of-the-art} shows the current situation of hardening existing systems.
Administrator Alice is responsible for a server running essential business functions (see \autoref{fig:TEST:state-of-the-art}, \textit{Scenario}).
In reality, these functions are automatic tests in different levels of abstraction from unit tests to end-to-end tests.

Alice wants to configure the server securely and uses a \cisShort{} guide.
This guide has more than $ 500 $ rules.
Alice automatically checks how many rules of the guide the system is currently not compliant with (see \autoref{fig:TEST:state-of-the-art}, \textit{Step 1}).

A recent study showed that a system using Windows 10 or Microsoft Office in the default configuration is, on average, only compliant with $ \approx 17.7 \% $ of the corresponding CIS rules~\cite{stockle2020automated}.
Thus, the checks will report more than 410 non-compliant rules.
Alice could go through these rules and make for each rule two decisions (see \autoref{fig:TEST:state-of-the-art}, \textit{Step 2a}):
First, is this rule important for the security of her server?
A specific rule might be beneficial in general, but the threat addressed by the rule might not be relevant for her server.
Second, could the rule interfere with the current behavior of the system, i.e., is the rule disabling some functionality the existing systems on the server still need?

Both decisions are complex and time-consuming.
For the first one, we need a threat analysis to identify the relevant assets and resulting threats.
For the second one, we need to know the potential side effects.
For each side effect, we then have to check whether it affects the software running on the server.
Although the rules include a description of the potential side effect, one must know how the existing software works to estimate potential problems with the rules to apply.
If we calculated with one minute per decision, the process would last more than six hours.
There are only a few systems where such an extensive analysis is economically reasonable.
Thus, Alice has two choices.
Either she applies all non-compliant rules on the server or does not harden the server.

If she chooses to apply all non-compliant rules, most probably problems in the system functionality will arise, and such problems will be revealed by re-running regression tests after the hardening process.
Thus, we assume for our example that there are broken functionalities.
If Alice knows the software, she can guess which rules might cause the problems (see \autoref{fig:TEST:state-of-the-art}, \textit{Step 2b});
we call these rules functionality-breaking or just \textbf{breaking} rules.
Otherwise, she has to disable rules until she finds all breaking rules.

\begin{figure*}[t]
    \includegraphics[width=\textwidth]{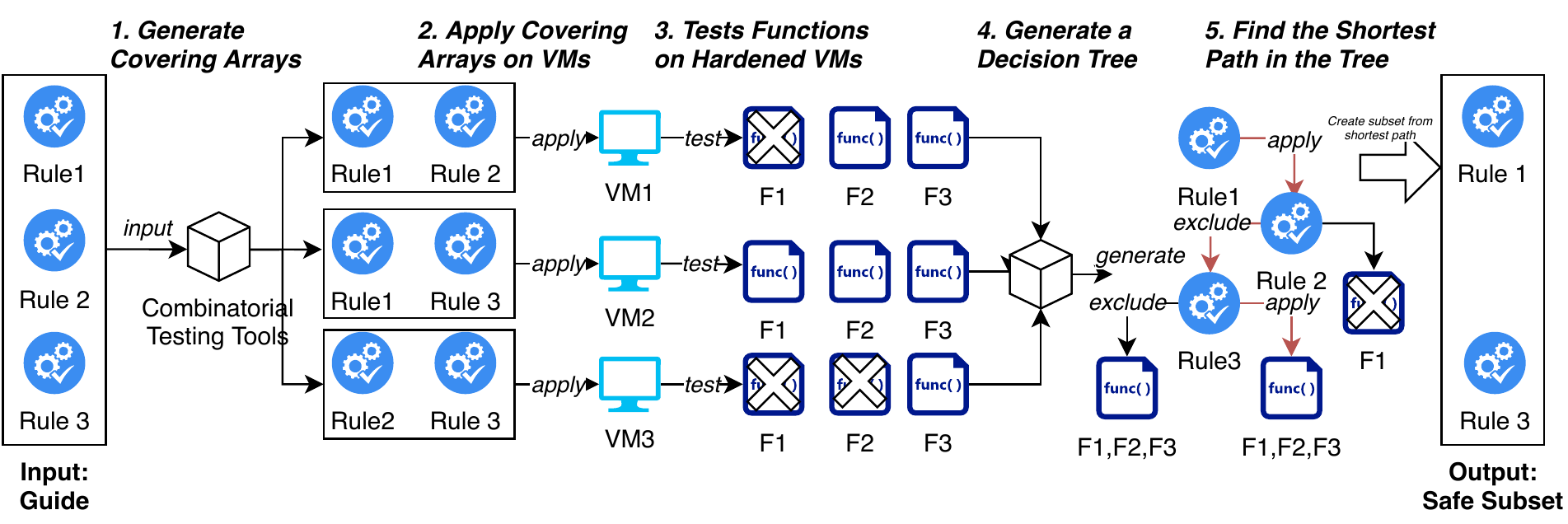}
    \caption{Identifying breaking rules using combinatorial testing and decision trees.}
    \label{fig:TEST:identifying-breaking-rules-using-combinatorial-testing-and-decision-trees}
\end{figure*}

If Alice found all the breaking rules and applied all other rules, she could guarantee the system's functionality and maximize the security gained by the configuration hardening.
Usually, Alice does not know how many rules she has to exclude and how the rules work together.
Multiple rules can break the same functionality, so she excludes all of them.
Furthermore, she might not have to remove all breaking rules, as combining multiple rules breaks one functionality, so she excludes one of the corresponding rules.

When we applied a CIS guide to a test system at Siemens, we had to exclude $ 9 $ of the $ 500 $ rules, i.e., if we knew the number of rules to exclude, we would \textit{only} have $ \binom{500}{9} $ candidates.
In practice, we do not need to investigate that many candidates, but, on average, this will cost far more time than the decision process, rendering this approach even more costly than the other.

Therefore, Alice will implement only rules for which she is 100\% sure that they do not break a function or not harden the system at all.
This behavior of dodging the risk of problems with the software functionality by neglecting the security configuration is widespread.
We conducted a case study with two municipalities in Southern Germany:
Although a guide existed for those municipalities, their systems only fulfilled 12\% and 35\% of the rules, respectively.
They argued that they have a very heterogeneous environment and high availability requirements and, thus, did not want to tamper with the system's functionality.
One might see this only as anecdotic evidence, but we heard the same argumentation from administrators at the Technical University of Munich.

\subsection{Problem and Proposed Solution}

The main problem addressed in this article is the following:
We want to harden an existing system without interfering with its current functionality.
Therefore, we want to find for a given guide, a given system and its given functionality a maximal subset of rules that does not break the functionality of the system.

To address this problem, we use existing combinatorial testing approaches combined with decision trees to efficiently find such a maximal subset (see \autoref{fig:TEST:identifying-breaking-rules-using-combinatorial-testing-and-decision-trees}).
First, we generate covering arrays based on the given rules.
Second, we apply the rules specified in the current array entry, test all the functions, and store the corresponding result.
Third, we train decision trees on the data from the previous step.
Fourth, we use the shortest path leading to all functions working to find a set of breaking rules that leads us to maximal subset that does not break the functionality.

Our contribution is that we transfer established techniques from the software engineering, or more specifically, the software testing domain, to the security configuration domain to solve a widespread problem.
Additionally, we share our code so practitioners can use it to find breaking rules.\footnote{\href{https://github.com/tum-i4/Better-Safe-Than-Sorry}{GitHub.com/tum-i4/Better-Safe-Than-Sorry}}
\subsection{Definitions and Concepts}

We use the motivating example to illustrate the definitions used in this article.
We denote systems in general with $ \xi $ and Alice's server with $ \xi_A $.
We denote rules as $ r \in \mathcal{R} $ and guides as set of rules, i.e., $ \mathcal{G} = \left\{ r_0, \dots, r_n \right\} $.
Furthermore, we abstract the functions as a set of predicates, i.e., $ \mathcal{F} = \left\{ f_1, \dots, f_n \right\} $ in general and the functions of Alice's server in particular as $ \mathcal{F}' $.

\begin{equation*}
    f\left( \xi , t \right) \Leftrightarrow \text{System } \xi \text{ fulfills the function } f \text{ at time } t
\end{equation*}

\noindent
Here and in following formulas, we use the time as parameter to differentiate between different states of a system $ \xi $.
For example $ f ( \xi, t ) $ and $ \neg f ( \xi, t + 1)$ means that $ \xi $ fulfills $ f $ at $ t $, but one step later, e.g., after apply some rules, not anymore.
Thus, the server $ \xi_A $ in our example has to fulfill $ \mathcal{F}' $.
Next, we denote the check with $ \aleph $. 

\begin{equation*}
    \aleph \left( \xi, t, r \right) \Leftrightarrow \text{System } \xi \text{ is compliant to rule } r \text{ at time } t
\end{equation*}

\noindent
Consecutively, we denote with the $ \mathcal{R}^{\mathit{NC}}_{\mathcal{G},\xi,t} $ all non-compliant rules of $ \mathcal{G} $ on system $ \xi $ at time $ t $.

\begin{equation*}
    \mathcal{R}^{\mathit{NC}}_{\mathcal{G},\xi,t} = \left\{ r | r \in \mathcal{G}: \neg \aleph \left(\xi, t, r \right) \right\}
\end{equation*}

\noindent
In our example case, we would have $ | \mathcal{R}^{\mathit{NC}}_{\mathcal{G}_{\mathit{CIS}}, \xi_A , t_{\mathit{Start}} } | \ge 410 $.
Next, we denote the broken functionalities of $ \mathcal{F} $ on the system $ \xi $ at time $ t $ with $\mathcal{F}^{F}_{\mathcal{F},\xi, t} $.

\begin{equation*}
    \mathcal{F}^{F}_{\mathcal{F},\xi, t} = \left\{ f | f \in \mathcal{F}: \neg f \left(\xi, t \right) \right\}
\end{equation*}

\noindent
Again, applied on our example, we would have $ | \mathcal{F}'^{F}_{\mathcal{F}',\xi_A, t_{\mathit{Start}}} | \ge 1 $ because there is a broken function.
Additionally, we denote with $ b $ that a set of rules $ \mathcal{R} $ is breaking, i.e., there is a failing functionality $ f $ after we applied all rules in $ \mathcal{R} $.

\begin{align*}
  b & \left( \xi, t, \mathcal{R}, \mathcal{F} \right) \Leftrightarrow ( ( ( \bigwedge_{f \in \mathcal{F}} f \left( \xi, t \right) ) \wedge ( \exists r \in \mathcal{R}: \neg \aleph \left(\xi, t, r \right)) \\
  & \wedge ( \bigwedge_{r \in \mathcal{R}}  \aleph \left( \xi, t + 1 , r \right) )) \Rightarrow \exists f \in \mathcal{F}: \neg f \left( \xi, t + 1 \right) )
\end{align*}

\noindent
We denote the breaking rules of $ \mathcal{G} $ on a system $ \xi $ at a time $ t $ with respect to the functionalities $ \mathcal{F} $ with $ \mathcal{B}_{\mathcal{G}, \xi, t, \mathcal{F} }$.

\begin{equation*}
    \mathcal{B}_{\mathcal{G}, \xi, t, \mathcal{F} } = \left\{ r | r \in \mathcal{G} : b \left(\xi, t, r, \mathcal{F} \right) \right\}
\end{equation*}

\noindent
Similarly, we define with $ n $ for a set of rules $ \mathcal{R} $ that they are \textbf{non-breaking} on a system $ \xi $ at a time $ t $ with respect to the functions $ \mathcal{F} $.

\begin{align*}
    n & \left( \xi, t, \mathcal{R}, \mathcal{F} \right) \Leftrightarrow \\
    & \left( \bigwedge_{f \in \mathcal{F}} f \left( \xi, t \right) \wedge \bigwedge_{r \in \mathcal{R}} \aleph \left(\xi, t + 1, r \right) \right) \Rightarrow  \bigwedge_{f \in \mathcal{F}} f \left(\xi, t + 1 \right)
\end{align*}

\noindent
Based on the definition of $ n $, we can now define a maximal non-breaking set:
A subset $ \mathcal{G} ^ *  $ is a maximal non-breaking set with respect to the system $ \xi $, at the time $ t $, and the functions $ \mathcal{F} $ if $ \mathcal{G} ^ *$ is non-breaking and any additional rule $ r' $ added to $  \mathcal{G} ^* $ would break at least one functionality.

\begin{align*}
    \mathcal{G} ^ * & \text{ maximal non-breaking with } \mathcal{G}, \mathcal{F}, t \Leftrightarrow \mathcal{G} ^ * \subset \mathcal{G} \\
    &  \wedge n \left( \xi, t, \mathcal{G} ^ *, \mathcal{F} \right) \wedge \forall r' \in \mathcal{G} \setminus \mathcal{G} ^ *: \neg n \left( \xi, t, \mathcal{G} ^ * \cup \{ r' \}, \mathcal{F}  \right)
\end{align*}

\noindent
Therefore, the goal of our approach is to find for a given system $ \xi $, given functions $ \mathcal{F}$, and a given guide $ \mathcal{G} $ at the time $ t $ a maximal non-breaking set $ \mathcal{G} ^ * $.

\section{Generate Covering Arrays From Security-Configuration Guides}
\label{sec:TEST:generate-covering-arrays-from-guides}

\begin{listing}[b]
\begin{minted}{ini}
[System]
Name: all_rules
[Parameter]
R1_1_1 (boolean) : true, false
R1_1_2 (boolean) : true, false
\end{minted}
\caption{CIS guide transformed into the \actsShort{} input format.}
\label{lst:TEST:acts-input}
\end{listing}

The naive approach to solving our problem would be to test every possible combination of rules $ \mathcal{P} \left( \mathcal{G} \right) $.
We could search the combinations without failing tests for the set with the most applied rules, but this is a very inefficient approach.
In the domain of software testing, researchers have already solved a similar problem:
If we want to test a program with many different parameters, we want to test it in all possible combinations of the parameters.
We can use combinatorial testing to test programs \textit{enough} without testing all the parameter combinations~\cite{kuhn2004software}.
Depending on the desired strength, we can drastically reduce the combinations to test with combinatorial testing compared with testing all combinations.
However, we can only reliably detect all faults up to this level, i.e., combinatorial testing of strength 2 can detect all faults caused by the combination of two or fewer parameters~\cite{kuhn2009combinatorial}.

If we apply the combinatorial testing approach, we need to decide for a targeted strength for finding breaking rules in a guide.
Since there is no data from the security-configuration domain, we have to rely on data from software testing.
A study by Kuhn et al. found no failing tests with a combination of more than six parameters~\cite{kuhn2004software}.
Thus, we \textit{assume} that there is \textbf{no} combination of \textbf{more than six rules} causing a function to break.
However, we discuss this strong assumption later in \autoref{sec:TEST:threats-to-validity}.

To apply combinatorial testing in the security-configuration domain, we generate once for every guide $ \mathcal{G} $ with $ n = | \mathcal{G} | $ rules a set of $ n $-tuples with \verb|true| or \verb|false|;
\verb|true| at the position $ i $ of a tuple means we apply the rule $ i $ in this combination.
In combinatorial testing, such a tuple is called \caLong{}.
We can reuse these \caShorts{} for multiple systems with different functionalities to find breaking rules.
If we add or remove new rules, we must regenerate the \caShorts{}.
Since the guides contain hundreds of rules, we needed algorithms that could handle tuples of that size.
Thus, we use the IPOG~\cite{lei2007ipog} and IPOG-D~\cite{lei2008ipog} algorithms and their implementations in the \actsLong{} (ACTS) tool to generate the combinations~\cite{yu2013acts}.

We first translate the rules of a guide from their original format into an \actsShort{} input file defining the used parameters.
In this input file, each parameter has a name and a data type, i.e., in our scenario, the rule's ID, e.g., \verb|R1_1_1|, and \texttt{boolean}.
One can see an example in \autoref{lst:TEST:acts-input}.
Depending on the chosen degree of \caShorts{} and algorithm, \actsShort{} now generates the \caShorts{}.
One can see an example output in \autoref{lst:TEST:acts-output}.
Afterward, we translate the \actsShort{} output into JSON files we use to implement guides automatically~\cite{stockle2020automated}.
We included several versions of these JSON files for IPOG and IPOG-D and different degrees in our repository.
Furthermore, one can use our published Python code to replicate these steps on their own.
After this first step, we now have the different \caShorts{} of the rules in our guide in a form we can automatically implement on a system to test tuple break the system's functionalities.

\begin{listing}[t]
\begin{minted}{text}
# Degree of interaction coverage: 2
# Number of parameters: 507
# Number of configurations: 20
R1_1_1,R1_1_2,R1_1_3,R1_1_4,R1_1_5,R1_1_6,...
true,true,true,true,false,false,...
true,true,false,false,true,true,...
false,false,true,true,true,true,...
false,false,false,false,false,false,...
...
\end{minted}
\caption{\actsShort{} output for a CIS guide.}
\label{lst:TEST:acts-output}
\end{listing}

\section{Tests Functions on Hardened Instances}
\label{sec:TEST:testing-functionalities-on-harden-systems}

\begin{listing}[b]
\begin{minted}{json}
[{ "name": "custom_1", "breaking": false,
   "rules": ["R1_1_1", "R1_1_2", "R1_1_3", "..."]},
{  "name": "custom_2", "breaking": true,
   "rules": ["R1_1_1", "R1_1_2", "R1_1_5", "..."]}, 
"..."]
\end{minted}
\caption{Example results of the testing process.}
\label{lst:TEST:results}
\end{listing}

Depending on the degree of the covering arrays, we generate between 20 and 5545 tuples for the CIS Windows 10 guide.
In the next step, we apply every tuple of rules and use the given tests $ \mathcal{F} $ to check if all functions are still accessible.
If a test fails, we record this in a log file.
After we have executed this procedure for every tuple, we collect the different log files and merge them.
The resulting file states which tuple broke a test and which did not.

What sounds straightforward, in theory, was cumbersome to realize.
The first problem was that we had to prepare an environment where we could set up the software, apply all rules in a tuple, test the functionalities, and record the result.
To solve this problem, security experts at Siemens use a toolchain with Ansible, Vagrant, and AWS instances to efficiently provision several virtual machine (VM) instances that they can configure independently and in parallel~\cite{stockle2022hardening}.
An alternative is to use Vagrant and a hypervisor like VirtualBox to run the VMs locally.
The second problem was ordering effects, i.e., a test is failing not because of the currently applied tuple but the previous one.
To avoid these effects, we could reset the VM after every test run or do a soft reset where we only revert the applied rules.
The hard reset costs more time than the soft reset, but there is no risk of side effects, e.g., if the revert mechanism of a rule does not work.
With the code in our repository, one can generate one VM instance for every tuple or several instances and distribute the tuples uniformly over the instances.
The third problem was the automatic tests.
Ideally, we would use tests from the industry to test real-world functionality.
However, as we stated before, only a few companies apply configuration hardening to their systems.
Those companies manually check whether their systems still work after the configuration hardening.
Thus, we needed to create our automatic tests ourselves.
Nevertheless, if all tests pass although the rules break the functionality, we will not detect this in our testing process, but only on the productive systems;
it is, therefore, crucial to find suitable tests~\cite{pretschner2015defect}.

After solving those issues, our testing procedure consists of the following steps:

\begin{listing}[t]
\begin{minted}{text}
instances/vagrant_0
├─ Vagrantfile               # Includes the used Vagrant box
├─ cis_windows_10_1909       # The guide to apply
│  ├── sfera_automation.json # Export JSON with all the 
│  │                         # rules and CA tuples.
│  └── sfera_automation.ps1  # Library to apply the guide
├─ setup.ps1                 # Script for the test process
└─ tests                     # Test directory
   ├── test.ps1              # Script coordinating the tests
   ├── tst_acc_cr.ps1        # Example test
   └── tst_conf.json         # Test configuration, e.g., 
                             # which tuples should we execute
\end{minted}
\caption{Example Vagrant directory.}
\label{lst:TEST:vagrant-directory}
\end{listing}

\begin{enumerate}
    \item Prepare the image of $ \xi $:
First, we prepare an image, i.e., a Vagrant box, with our software and all needed dependencies as reference to set up the different instances.
    \item Prepare the instances $ \xi_i $:
We prepare for every VM instance $ \xi_i$ to start a directory with the software to run, the tests, the guide, and the \caShort{} tuples to test.
If we want to test several tuples on one instance, we distribute the tuples uniformly.
One can see the layout of such a directory in \autoref{lst:TEST:vagrant-directory}.
On the one side, we can fasten the instance setup if we add more actions into the box setup.
On the other side, we want the image to be flexible and fit more than one system, thus keeping it slim.
    \item Start the instances $ \xi_i $:
We start the instances either in parallel or sequentially.
    \item Is the system working?:
We execute all tests $ \mathcal{F} $ in one instance $ \xi_0 $ to see whether the functionality works on an instance in the default configuration.
If the tests fail before applying any rules, i.e., $ | \mathcal{F}^{F}_{\mathcal{F},\xi_0, t_0} | \ge 1 $, there must be a problem in the tests or the setup that we need to fix.
    \item Apply all rules in $ \mathcal{G} $:
We apply all rules on one instance $\xi_0 $, i.e., $ \mathcal{R}^{\mathit{NC}}_{\mathcal{G}, \xi_0 ,t_1} = \emptyset $
\item Run all tests in $ \mathcal{F} $:
If there is a breaking rule in the guide, we will see at least one test failing, i.e., $ | \mathcal{F}^{F}_{\mathcal{F},\xi_0, t_1} | \ge 1 $.
If no tests fail, i.e., $ \mathcal{F}^{F}_{\mathcal{F},\xi_0, t_1}  = \emptyset $  we can safely apply all rules and stop the testing process at this point.
    \item Revert all rules:
If we use the soft reset, the revert mechanism must work, i.e., $ \mathcal{R}^{\mathit{NC}}_{\mathcal{G},\xi_0,t_2} = \mathcal{R}^{\mathit{NC}}_{\mathcal{G},\xi_0,t_0} $.
However, there are some rules that we cannot revert.
Thus, we try to revert all rules and execute the tests again.
If some tests still fail, i.e., $ | \mathcal{F}^{F}_{\mathcal{F},\xi_0, t_2} | \ge 1 $, there is a problem with the revert mechanism.
    \item Apply a tuple:
We take the first tuple of the list of untested tuples and apply all rules $ \mathcal{G}_j \subset \mathcal{G} $ corresponding to this tuple on an instance $ \xi_i $, i.e., $ \mathcal{R}^{\mathit{NC}}_{\mathcal{G},\xi_i, t'_j} = \mathcal{R}^{\mathit{NC}}_{\mathcal{G},\xi_i, t_0} \setminus \mathcal{G}_j $ with $ t' $ being after the application.
    \item Test the functionality $ \mathcal{F} $ :
We execute the tests and store whether there were failing tests, i.e., $ | \mathcal{F}^{F}_{\mathcal{F}, \xi_i, t'_j} | \ge 1 $, in a JSON file.
    \item Reset:
To prepare a clean environment again, we perform a reset so that $ \mathcal{R}^{NC}_{\mathcal{G},\xi_i,t''} = \mathcal{R}^{NC}_{\mathcal{G},\xi_i,t_0} $.
Afterward, we go back to Step~7 until we have applied all tuples.
    \item Collect the results:
After we have applied and tested all tuples, we collect all results, i.e., $ \mathcal{G}_j $ and whether there were failing tests $ | \mathcal{F}^{F}_{\mathcal{F}, \xi_i, t'_j} | \ge 1 $, from the different instances $ \xi_i $ and combine them in a single JSON file.
One can see an example output in \autoref{lst:TEST:results}.
    \item Tear down:
We destroy the instances $ \xi_i $.
\end{enumerate}

Again, one can use our published code to redo these steps on their own.
We have now tested all tuples, and the resulting JSON states which tuples cause which tests to fail.
In the next step, we use this information to deduce which rules caused the failures.

\section{Analyze the Test Results}
\label{sec:TEST:analyze-the-test-results}

\begin{figure}[t]
\centering
\includegraphics[width=\linewidth]{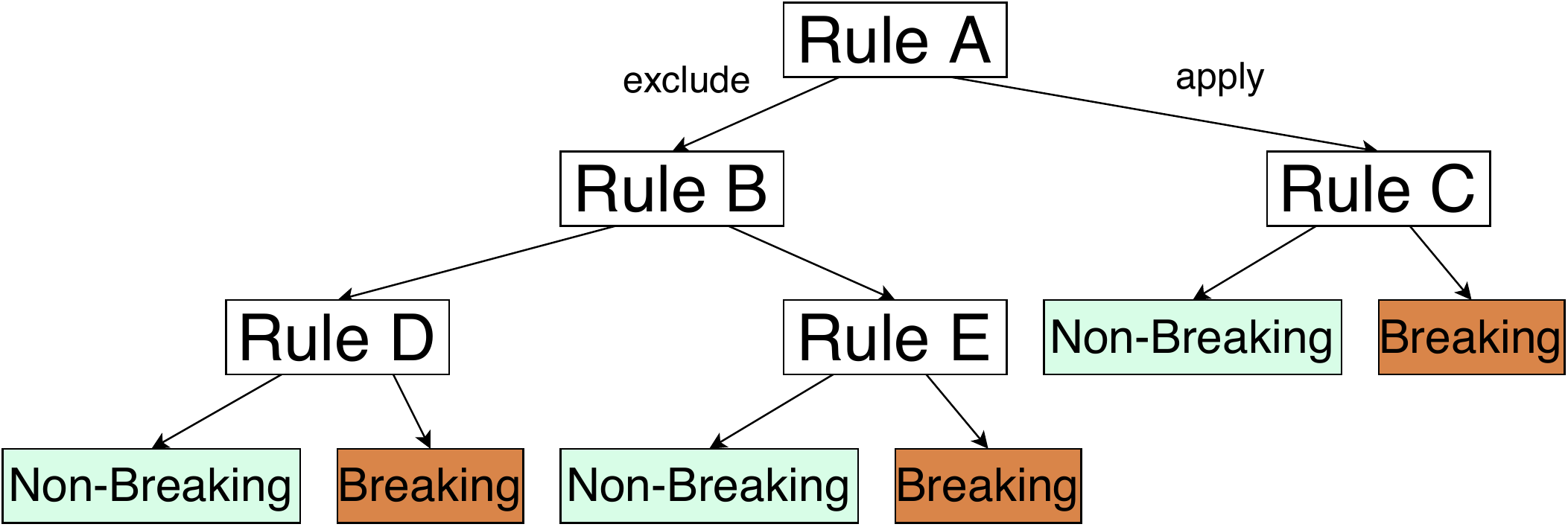}
\caption{Simplified example of a generated decision tree.}
\label{fig:TEST:approach-decision-tree-example}
\end{figure}

As a result of the previous step, we know for each of the tuples whether the application of these rules broke the functionality or not (see \autoref{lst:TEST:results}).
Next, we analyze these data to find a maximal non-breaking set $ \mathcal{G} ^ * \subset \mathcal{G} $.
We could pass the failing tuples to the administrators like Alice so that they adjust the software on $ \xi $ to work even when these tuples are applied.
However, especially for legacy systems, it may be challenging to carry out changes;
Therefore, we want to adjust the guide by removing potential problematic rules for our general scenario.
At Siemens, the administrators would decide whether they need other security measures to reduce the risks resulting from the excluded rules.

We can visualize the results of the previous step in a truth table.
Every row is a \caShort{} tuple $ \mathcal{G}_j $ , every column is a rule $ r \in \mathcal{G} $, and the last column is the result of the tests, i.e., $ \mathcal{F}^{F}_{\mathcal{F}, \xi_i, t'_j} \ge 1$.
Finding a minimal cutting set in such a structure is a common task in different computer science disciplines, and there are different solutions.
We used two different approaches for our proof of concept.

\subsection{Decision Trees}
\label{sub:TEST:decision-trees}

First, we adapted the approach described by Yilmaz et al.~ \cite{yilmaz2006covering} for our scenario.
They used machine learning to find the parameters causing a test to fail and trained a decision tree on their test results.
Thus, we ported their approach into our domain and learned decision trees on our results.

One can see an example decision tree in \autoref{fig:TEST:approach-decision-tree-example}.
The nodes in the tree represent rules from the tuples.
The algorithm calculated different partitions of breaking and non-breaking tuples by checking whether we apply a rule $ r $ in $ \mathcal{G}_j $, i.e., $ r \in \mathcal{G}_j $, or not.
In \autoref{fig:TEST:approach-decision-tree-example}, the algorithm differentiates first between tuples where \textit{Rule A} is applied or not.
Every leaf node states whether the functionality was broken, i.e., $ | \mathcal{F}^{F}_{\mathcal{F}, \xi_i, t'_j} | \ge 1$, or not when we applied or excluded the rules on the path between the root and the leaf node.
In \autoref{fig:TEST:approach-decision-tree-example}, applying \textit{Rule A} but excluding \textit{Rule C} leads to a non-breaking leaf, i.e., \textit{Rule C} is a breaking rule.

\begin{figure}[t]
\centering
\includegraphics[width=0.7\linewidth]{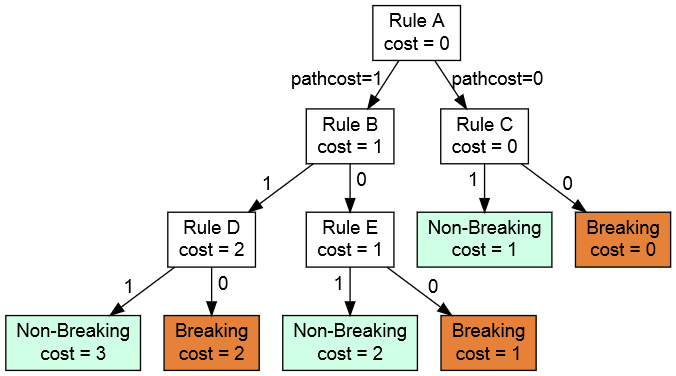}
\caption{Visualization of the modified decision tree from \autoref{fig:TEST:approach-decision-tree-example} with weights to support shortest path search.}
\label{fig:TEST:approach-shortest-path-example}
\end{figure}

We could use this decision tree to predict whether a tuple that we have not tested before is breaking or not.
However, we are only interested in the non-breaking leaves with the least excluded rules.
Therefore, we use shortest path search algorithms like Dijkstra's algorithm on the decision trees to find these leaves.
We give edges that apply a rule, i.e., right edges in \autoref{fig:TEST:approach-decision-tree-example}, a value $0$, and all other edges the value $ 1 $.
One can see an example in \autoref{fig:TEST:approach-shortest-path-example}.
Therefore, the total cost of a path from the root to a leaf is the number of rules that we did not apply.
In turn, the shortest path leading to a non-breaking leaf is the path with the least number of rules not being applied, i.e., a maximal non-breaking set $ \mathcal{G} ^ * $.
In \autoref{fig:TEST:approach-shortest-path-example}, this is $ \mathcal{G} ^ * = \mathcal{G} \setminus \left\{ R_C \right\} $.

We implemented our approach using \emph{scikit-learn}~\cite{pedregosa2011scikit}.
First, we translate the data from the previous step into the scikit format.
Second, we train a decision tree on it.
Third, we added the weights to the learned decision tree.
Fourth, we run a shortest path algorithm on the weighted decision tree to find a non-breaking leaf.
Five, we follow the path from the root to this leaf.
If the current edge has the weight $ 1 $, we remove the current rule from the guide $ \mathcal{G} $.
The resulting set is a maximal non-breaking set $ \mathcal{G} ^ *  $.

\subsection{Logic Minimization}
\label{sub:TEST:logic-minimization}

In addition to the heuristic approach, we can formally examine the results with logic minimization to derive the minimal set.
Here, we pass our truth table representing our test results to one of the many minimization algorithms~\cite{brayton1984logic} and use the minimized table to find the maximal non-breaking set $ \mathcal{G} ^ * $.

We first transform the test results into a truth table in our implementation.
Second, we pass this table to the minimization library \emph{PyEDA}~\cite{drake2015pyeda}.
Third, we exclude the rules according to the minimized table to derive $ \mathcal{G} ^ * $.

We provided sample results and code in our repository so that one can redo these steps with both approaches.
In the end, we have a maximal non-breaking set $\mathcal{G} ^ * \subset \mathcal{G} $, and we can use it to harden our system safely without breaking any legacy functionality.

\section{Evaluation}
\label{sec:TEST:evaluation}

\begin{listing}[b]
\centering
\begin{minted}{json}
[["R1", "R2"], ["R3", "R4"]]
\end{minted}
\caption{Example definition of a breaking combination.}
\label{lst:TEST:breaking-rules}
\end{listing}

When evaluating our approach and its implementation, we focus on the following research questions:

\begin{description}
    \item[RQ1] Can we find a maximal non-breaking subset $ \mathcal{G} ^ * $ of a guide $ \mathcal{G} $ with respect to given functionalities $ \mathcal{F} $ using combinatorial testing?
    What degrees of breaking rules can we detect?
    In theory, the strength of the covering array should be the upper limit for the degree of breaking rules we can detect.
    However, the combination with the decision trees could reduce our approach's power in practice.
    \item[RQ2] How much time does our approach need?
    Is this a reasonable time effort for hardening a system?
\end{description}

We used the \cisShort{} guide for Windows~10 version 1909~\cite{security2022cis} for our evaluation.
This guide contains 507 rules, i.e., $ | \mathcal{G}_{\mathit{CIS},\mathit{W10}} | = 507 $.
To apply and revert the rules automatically, we transformed the guide into the Scapolite format~\cite{stockle2020automated}.
In the following, we discuss how we evaluated the different steps of our approach.

\subsection{Evaluation of the Covering Array Generation}

\begin{table}[t]
\caption{Number of generated tuples depending on the used algorithm and strength.}
\label{tab:number-cov-arrays}
\centering
\begin{tabular}{l r r r r}
    \toprule
    \diagbox{Algorithm}{Strength}    & 2 & 3 & 4 & 5 \\
    \midrule
    IPOG & 20 & 70 & 209 & - \\
    \midrule
    IPOG-D & 20 & 78 & 305 & 5545 \\
    \bottomrule
\end{tabular}
\end{table}

First, we had to evaluate the generation of the \caShorts{} from an existing guide.
As discussed in \autoref{sec:TEST:generate-covering-arrays-from-guides}, we wanted to generate \caShorts{} with strength from $2$ to $6$ with both IPOG and IPOG-D.
We evaluated how long the generation takes for the $ \mathcal{G}_{\mathit{CIS}, \mathit{W10}}$ guide to partially answer \textbf{RQ1}.

We run the ACTS tool on a server with two Intel Xeon E5-2687W v3 CPUs with 40 cores and a total of $500$ GB of RAM, from which we used up to $340$ GB.

\subsection{Testing Process}

We evaluated our testing process in three steps.

\subsubsection{Simulation}

In the first step, we simulated the breaking rules.
Here, we first defined combinations of rules that break the functionality.
We defined $ 51 $ combinations of breaking rules;
one combination is the empty set, i.e., the special case where we can apply our guide without problems.
We define them as logical formulas in disjunctive normal form;
\autoref{lst:TEST:breaking-rules} shows how we express $ \left( R1 \wedge R2 \right) \vee \left( R3 \wedge R4 \right) $.
If we apply all rules of the first or the second subformula, the system's functionality will break, e.g., we can apply \verb|R1| and \verb|R2| to break the function, but not \verb|R1| and \verb|R3|.
Furthermore, we tested for each of the non-empty combinations $ 3 $ additional random variants, i.e., we replaced the rules' ids with random other ids to avoid potential side effects based on the order of the rules.
In total, we thus tested $201$ sets;
one can find all $ 201 $ sets in our repository to reproduce our evaluation.

For each of the combinations, we then went through the list of tuples and created for each tuple the result file:
If a \caShort{} tuple includes a breaking combination, we mark the tuple as breaking else as non-breaking.
Afterward, we combine all result files and analyze them using our decision-tree-based and logic minimization approach.
We used different \caShorts{} from both generation algorithms and with different strengths.

\subsubsection{Validation of the Simulation}

In the second step, we evaluated the complete process but used generated mock tests based on the defined combinations from the Evaluation~Step~1.
As described in \autoref{sec:TEST:testing-functionalities-on-harden-systems}, we execute all steps.
However, the tests in Step~8 check for a given definition of breaking rules and whether the current system $ \xi $ is compliant with those rules.
If so, we mark the current tuple as failing.
We could test many combinations of breaking rules with those generated mock tests.
In contrast to the simulation, this step required significantly more time.
While executing the tests is cheap, setting up the VM and then applying and reverting the rules takes some time.
We conducted this part of the evaluation with the covering arrays of strength $ 4 $ generated by the IPOG-D on two VMs.

\subsubsection{Practical Application}

In the third step, we evaluated the process with a real test.
Here, we needed a small program $ f' $ that fails when the whole guide is applied, i.e., $ \mathcal{R}^{\mathit{NC}}_{\mathcal{G_{\mathit{CIS},\mathit{W10}}}, \xi, t'} = \emptyset $.
We chose a simple PowerShell script that creates a new user with a password and then deletes the user again.
This test symbolizes the nightmare of all administrators like Alice:
It is a straightforward task that usually takes seconds but does not work after the hardening.
Thus, they must spend hours investigating which rule broke the functionality that does not relate to the script's original time effort.
Next, we did all the steps of our process with two instances running in parallel and marked the tuples as failing when $ f' $ failed.
Again, we gave the result to our analyzing components to find $ \mathcal{G}_{\mathit{CIS},\mathit{W10}} ^ * $.
In the end, we applied $ \mathcal{G}_{\mathit{CIS},\mathit{W10}} ^ * $ and checked whether $ f'$ was working or not.
We again conducted this part of the evaluation with the covering arrays of strength $ 4 $ generated by the IPOG-D on two VMs.

The simulation helps us to answer \textbf{RQ1}.
In Step~2 and Step~3, we measured the time needed since this contributes significantly to the answer of \textbf{RQ2}.

\subsection{Evaluation of the Analysis}

\begin{table}[t]
    \caption{Time (in seconds) needed to generate covering arrays of given strength depending on the used algorithm.}
    \label{tab:time-cov-arrays}
    \centering
    \begin{tabular}{l r r r r}
        \toprule
        \diagbox{Algorithm}{Strength} & 2 & 3 & 4 & 5 \\
        \midrule
        IPOG & 0.7 & 374 & 179149 & - \\
        \midrule
        IPOG-D & 0.2 & 16 & 8451 & 1184478 \\
        \bottomrule
    \end{tabular}
\end{table}

We need the analysis of the test results to assess the general quality of our approach.
However, we also compared different factors:
First, we compared the analysis variant, i.e., the decision-tree-based heuristic and the logic minimization approach.
Second, we compared the influence of different parameters, e.g., for the decision tree generation.
Third, we compared different algorithms to find a solution in the tree-based approach, e.g., taking the non-breaking leaf with the most samples.

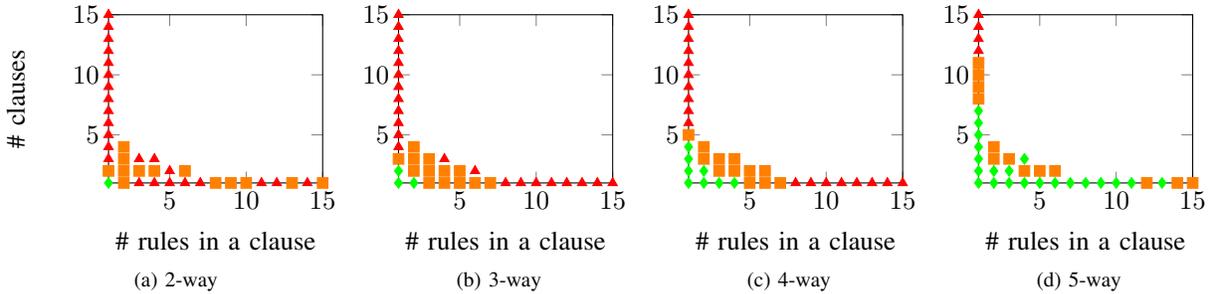
\begin{figure*}[t]
    \centering
\subfloat[2-way]{%
\begin{tikzpicture}
    \begin{axis}[
        width =0.5\columnwidth,
        enlargelimits = false,
        xlabel = {\# rules in a clause},
        ylabel = {\# clauses},
    ]
    \addplot+[
        only marks,
        scatter,
        mark = *,
        point meta = explicit symbolic,
        scatter/classes={
            0={mark=triangle*,red},
            1={mark=square*,orange},
            2={mark=diamond*, green}
        }
    ]
    table[meta=2]{data/breaking_rules_sets.dat};
    \end{axis}
\end{tikzpicture}
}
\subfloat[3-way]{%
    \begin{tikzpicture}
        \begin{axis}[
            width =0.5\columnwidth,
            enlargelimits = false,
            xlabel = {\# rules in a clause},
        ]
        \addplot+[
            only marks,
            scatter,
            mark = *,
            point meta = explicit symbolic,
            scatter/classes={
                0={mark=triangle*,red},
                1={mark=square*,orange},
                2={mark=diamond*, green}
            }
        ]
        table[meta=3]{data/breaking_rules_sets.dat};
        \end{axis}
    \end{tikzpicture}
}
\subfloat[4-way]{%
    \begin{tikzpicture}
        \begin{axis}[
            width =0.5\columnwidth,
            enlargelimits = false,
            xlabel = {\# rules in a clause},
        ]
        \addplot+[
            only marks,
            scatter,
            mark = *,
            point meta = explicit symbolic,
            scatter/classes={
                0={mark=triangle*,red},
                1={mark=square*,orange},
                2={mark=diamond*, green}
            }
        ]
        table[meta=4]{data/breaking_rules_sets.dat};
        \end{axis}
    \end{tikzpicture}
}
\subfloat[5-way]{%
    \begin{tikzpicture}
        \begin{axis}[
            width =0.5\columnwidth,
            enlargelimits = false,
            xlabel = {\# rules in a clause},
        ]
        \addplot+[
            only marks,
            scatter,
            mark = *,
            point meta = explicit symbolic,
            scatter/classes={
                0={mark=triangle*,red},
                1={mark=square*,orange},
                2={mark=diamond*, green}
            }
        ]
        table[meta=5]{data/breaking_rules_sets.dat};
        \end{axis}
    \end{tikzpicture}
    }
\caption{Distribution of the breaking rules sets.}
\label{fig:TEST:breaking-rules-sets}
\end{figure*}

We assessed for the different variants whether they calculated a correct maximal non-breaking set $ \mathcal{G} ^ *$ and checked how much they differ in the case of incorrect output.
Thus, we can answer \textbf{RQ1}.
Moreover, we measured how long the calculation of the breaking rules takes depending on the input, contributing to the answer of \textbf{RQ2}.

\section{Results}
\label{sec:TEST:results}

In this chapter, we will present the results of the evaluation of our proof of concept.

\subsection{Results of the Covering Array Generation}

\autoref{tab:number-cov-arrays} shows the number of tuples per covering array for different strengths and both IPOG and IPOG-D.
The number of tuples grows exponentially with increasing strength.
However, due to the required time, we could not generate covering arrays of the strength of $ 5 $ using IPOG or $ 6 $ with IPOG-D.
\autoref{tab:time-cov-arrays} shows the required time to generate the covering arrays.
For all strengths between $2$ and $4$, the time is an average of $5$ measurements;
for $ 5 $, we measured the time only once.
One can see that IPOG-D -- as expected -- is faster than IPOG.
Thus, for IPOG-D, we could even calculate the covering arrays with strength $ 5 $, although it took $ \approx 13.7$ days.
However, IPOG-D generated more tuples for covering arrays of the same strength.

\subsection{Results of the Testing Process}

\begin{listing}[b]
\begin{minted}{json}
[["R2_2_1", "R2_2_2", "R2_2_3"],
 ["R2_2_4", "R2_2_5", "R2_2_6"],
 ["R2_2_7", "R2_2_8", "R2_2_9"]]
\end{minted}
\caption{Problematic breaking rule set.}
\label{lst:three-three-one}
\end{listing}

\subsubsection{Simulation}

\autoref{fig:TEST:breaking-rules-sets} shows our simulation results with covering arrays of strength 2 to 5.
We cluster the breaking rule set based on the number of clauses they have on the x-axis and what is the maximal number of rules within a clause on the y-axis.
A red triangle means that our approach could not identify a single solution correctly for all sets in this cluster.
An orange square means that our approach identified some solutions correctly for the sets in this cluster.
A green diamond means that our approach identified all solutions correctly for this cluster.
Based on the strength of the covering arrays, we expected a green triangle in the lower left corner.
However, our approach performed worse than expected for 2-way and 3-way arrays but better for 5-way arrays.
The most prominent reason is the lack of data for the decision tree.
The decision tree has no data to partition and, therefore, cannot determine any rules to be excluded.
In some cases, the algorithm could find subsets of the correct solutions.
We consider these cases invalid results for this evaluation, but they may contribute to finding the optimal solution.
Surprisingly, our approach could calculate the correct results for single clauses with up to $11$ rules per clause based on the covering arrays of strength $ 5 $, although we only expected correct results for up to $5$.
We investigated the breaking rule sets that were not calculated correctly in more detail.
Our approach could not determine the correct solution for combination in \autoref{lst:three-three-one} and all its variants.
Instead of excluding one rule from each of the three clauses, our algorithm only excludes \verb|R2_2_4| and \verb|R2_2_7|, i.e., a subset of one correct solution.
Thus, we investigated the corresponding decision tree.
Although correct solutions were part of the tree, the wrong solutions dominated them because of the shorter length.
We are not sure why the decision tree was partitioned, but maybe more test tuples would have helped.
In general, our approach calculated the correct solution $ 77\%$ of the breaking rules in our sample set.
One could argue that this is too low to use the approach in practice, but our sample set includes far more complicated combinations than we expect in reality.
On the realistic samples, based on the assumption that only up to 6 rules combined lead to a problem, we achieve almost 100\%.

\subsubsection{Validation of the Simulation}

When we started this part of the evaluation, the initial tests failed, i.e., Step~4 of the testing process.
These initial failing tests never happened for the simulation.
The reason is those mentioned above $ \approx 17.7\% $ compliant rules on a system in its default configuration.
If all rules in a clause are already compliant with a default system, the mock test fails at Step~4.
Thus, we had to skip those tests;
an alternative was to use an image non-compliant with all rules, but we deemed this not a realistic scenario.
Apart from this, the generated mock tests lead to the same data as the simulation.
The whole process took around $12$ hours.

\subsubsection{Practical Application}
\label{subsub:TEST:result:practical-application}

\begin{figure}[t]
\centering
\includegraphics[width=0.65\linewidth]{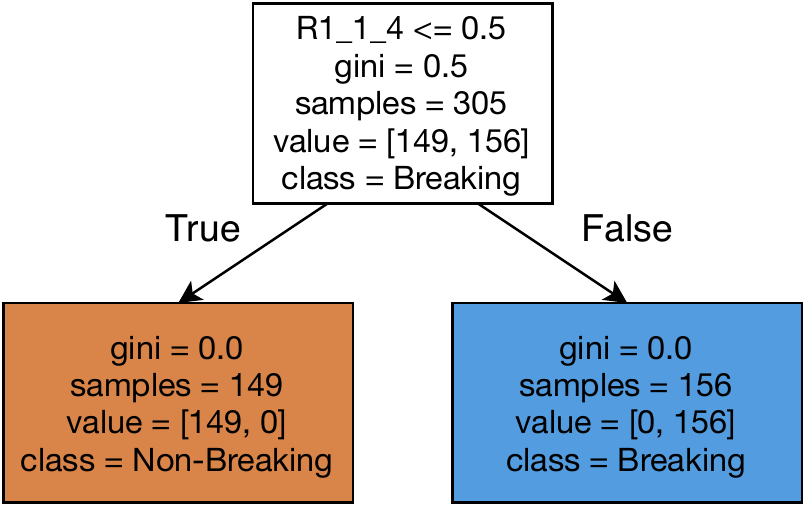}
\caption{Generated decision tree for the example functionality.}
\label{fig:TEST:decision-tree-vagrant}
\end{figure}

We evaluated the practical application of our proof of concept by testing functionality with unknown breaking rules.
We knew from the initial tests that the functionality worked before applying any rules but did not work after applying them.
Overall, the testing process took 12 hours to complete, resulting in $156$ breaking and $149$ non-breaking combinations.
\autoref{fig:TEST:decision-tree-vagrant} shows the decision trees learned on these results stating that one should exclude rule \verb|R1_1_4|.
Afterward, we applied $ \mathcal{G} ^ * = \mathcal{G}_{\mathit{CIS},\mathit{W10}} \setminus \left\{ R1\_1\_4 \right\} $, ran our test again, and it succeeded.
Rule \verb|R1_1_4| ``ensure [that] `Minimum password length' is set to `14 or more character(s)' '', but our test script tried to set a password of length $ 6 $ and thus failed.

\subsection{Evaluation of the Analysis}

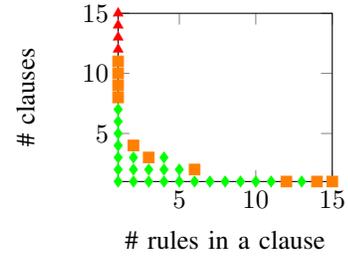
\begin{figure}[t]
    \centering
    \begin{tikzpicture}
        \begin{axis}[
            width =0.5\columnwidth,
            enlargelimits = false,
            xlabel = {\# rules in a clause},
            ylabel = {\# clauses},
        ]
        \addplot+[
            only marks,
            scatter,
            mark = *,
            point meta = explicit symbolic,
            scatter/classes={
                0={mark=triangle*,red},
                1={mark=square*,orange},
                2={mark=diamond*, green}
            }
        ]
        table[meta=m]{data/breaking_rules_sets.dat};
        \end{axis}
    \end{tikzpicture}
    \caption{Distribution of the breaking rules sets when using combinations of strength 5 and the modified approach.}
    \label{fig:breaking_rules_sets_5-way_modified}
\end{figure}

Next, we compared different techniques to find the optimal solution.
First, we select the non-breaking leaf with the biggest partition instead of the leaf with the shortest path.
We developed this potential improvement based on the problems seen, e.g., in combinations like \autoref{lst:three-three-one}.
We used the 5-way combinations generated by IPOG-D to compare the different path algorithms.
\autoref{fig:breaking_rules_sets_5-way_modified} shows the result of the modified approach.
One can see that the figure is almost identical to \autoref{fig:TEST:breaking-rules-sets}, but the modified approach performed better in the mixed clusters like \autoref{lst:three-three-one}, i.e., some orange triangles are here green diamonds.

Apart from modifying the used path algorithm, we also evaluated modifications of the generation of the decision tree, e.g., introducing a minimum for the number of test cases in a partition or changing the minimum number of data points for a split of the partition.
However, none of these changes lead to identifying more correct solutions.

We also evaluated the logic optimization approach compared to the decision tree-based heuristic approach.
The logical optimization took more time than the heuristic.
Thus, we excluded breaking rules sets with more than $5$ clauses, more than $9$ rules per clause, and did not evaluate the random variants of breaking rules sets.
Since more test cases increased the size of the truth table to be minimized, calculation time heavily depended on the strength of the covering arrays;
the algorithm calculated the results of strength $ 2 $ and $ 3 $ in seconds, $ 4$ in minutes, but for $ 5 $, it took more than 24 hours.
Also, more clauses lead to longer computation.
As expected, the logical optimization calculated all solutions correctly within the strength of the covering arrays and most of the remaining solutions;
it incorrectly calculated only $3$ of the $35$.
The decision tree-based heuristic led to $ 4 $ incorrect results in this sample set.
Thus, the performance of the logic minimization approach is similar to the decision-tree-based heuristic.

\section{Discussion}
\label{sub:TEST:discussion}

In this chapter, we will discuss the results of the evaluation and answer our research questions.

\subsection{Finding Breaking Rules}
\label{sub:TEST:discussion:finding-breaking-rules}

The results of our evaluation showed that one can use combinatorial testing and specifically covering arrays to find breaking combinations of rules.
Furthermore, we can use logical minimization or machine-learning-based heuristics to find a maximal non-breaking set $ \mathcal{G} ^ * $.
However, our results also showed that there are some caveats.
First, we could not create covering arrays with a strength of 6, although we assumed this was the necessary strength to cover all breaking sets in practice.
Even generating covering arrays of a strength of 5 occupied too much memory (340GB) for (13 days) so this is hardly useful for public guides, e.g., from the CIS.
For private guides, e.g., at Siemens, it is not economical to spend that many resources on every guide.
Thus, it is more realistic to use 4-way covering arrays, guaranteeing that we will find all combinations of 4 breaking rules or less.
Nevertheless, the results show that our heuristic approach can find some solutions for higher combinations or at least subsets of an optimal solution that could help the administrators.
Since we have no information about the distribution of the breaking rules in practice, the answer of \textbf{RQ1} is twofold.
If all or most breaking functionality results in practice from 4 or fewer rules, covering arrays and our heuristic analysis can reliably identify the breaking rules.
If a significant portion of breaking functionality results from 5 or more rules, covering arrays and heuristic analysis can only help to identify the optimal solution.

\subsection{Effort}
\label{sub:TEST:discussion:effort}

Generating different covering arrays depends on the number of rules in the chosen guide and the desired strength of the combinations.
Based on that, it requires a couple of hours, days, or even weeks.
However, we need this generation only once when the publisher publishes the guide and not for every system we want to harden with the given guide.
The CIS also updates its guides, but if the rules change, they do not have to regenerate the covering arrays.
If they add new rules, they could reuse the existing arrays to speed up the generation.
As stated above, we would argue that it is more realistic to use 4-way covering arrays as 5-way covering arrays are too expensive.

We can estimate the effort of identifying breaking rules for a given system $ \xi $ using the following formula:

\begin{equation*}
    t_{\Sigma} = N_{\mathit{VMs}} \cdot t_{\mathit{VM}} + t_{\mathit{SW}} + \frac{N_{C}}{N_{\mathit{VMs}}} \left( t_{A}  + t_{T} + t_{\mathit{SR}} \right) + t_{\mathit{ANA}}
\end{equation*}

with
\begin{description}
    \item[$  N_{C} $] the number of tuples, e.g., 305 for the $ \mathcal{G}_{\mathit{CIS},\mathit{W10}} $ guide and 4-way covering arrays of the IPOG-D.
    \item[$  N_{\mathit{VMs}} $] the number of instances, e.g., 2 in our evaluation.
    \item[$  t_{\mathit{VM}}  $] the time needed to prepare the instances, i.e., Testing process, Step~2.
    \item[$  t_{SW}  $] the time to start the instances and set up the software whose functionality we want to ensure, i.e., Step~3.
    \item[$  t_{A} $] the time needed to apply all rules in a tuple, i.e., Step~8.
    \item[$  t_{T}  $] the time needed to execute the automatic tests, i.e., Step~9
    \item[$  t_{\mathit{SR}}  $] the time needed to do a soft reset of the applied rules, i.e., Step~10
    \item[$  t_{\mathit{ANA}} $] the time needed for the analysis of the test runs
\end{description}

Setting up a local VM, e.g., with Vagrant, can last for several minutes, whereas a VM in the cloud, e.g., on AWS, is much faster.
$ t_{SW} $ depends on the complexity of the software.
In our evaluation, we tested a core function of Windows and, thus, did not install any additional software.
In our previous study~\cite{stockle2020automated}, we showed that $ t_{A} $ is for Windows rules below one 1s per rule but not negligible when executed many times.
$ t_T $ depends on the complexity of the tests.
The script took a couple of seconds in our evaluation, but if one uses complex tests, this could take minutes or hours.
$ t_{\mathit{SR}} $ in contrast, is again in the order $ t_{A}$.
$ t_{\mathit{ANA}} $ with the decision trees, and the shortest path algorithm takes only a couple of seconds using our heuristic and several minutes with the logic minimization.

As mentioned in \autoref{subsub:TEST:result:practical-application}, our experiments run for more than 12h with covering arrays of strength of 4, answering partially \textbf{RQ2}.
This time might be reasonable if we execute it only for releases, but it is too much to use in continuous integration contexts.
We can mainly influence two factors in the equation: the number of tuples $  N_{C} $ and the number of instances $ N_{I} $.
If we reduce the number of tuples by choosing covering arrays with lower power, we will not detect some complex combinations.
Thus, we will increase $ N_{I} $ by using more instances in parallel, e.g., on-demand in the cloud.
If we used 30 on-demand instances, we only ran $ 11 $ combinations on every machine.
On average, a combination applied 240 rules, i.e., the application and software reset would last at most 240s.
Assuming that the automatic tests last 2min, the whole process would need 2h.
To estimate this process's price, we calculated 30 on-demand Windows instances, each with 2 cores, 4GB RAM, and 25GB storage on AWS.
The estimation there was that this would cost around \$3.
We argue that 2h is short enough to include this process into regression tests running every night and \$3 is cheap enough to execute the process for most systems.

\section{Threats to validity}
\label{sec:TEST:threats-to-validity}

In this section, we list potential threats to the validity of our results.

\subsection{Internal validity}

\subsubsection{Choice of the Breaking Rule Sets}

We evaluated our approach with our breaking rule sets.
We tried to cover a variety of possible breaking rules and added, e.g., formulas with no overlap, such that the result must contain one rule from each subformula.
However, there could still be an unintended bias in these sets.

\subsection{External validity}

\subsubsection{Lack of Good and Automated Tests}

One core assumption of this article is that there are automatic tests whose failure indicates that some function is broken.
However, in most organizations, there are no tests for their legacy systems at all.
Some organizations test their legacy systems manually, but even if the needed effort is minimal, repeating a manual task in combination with the covering arrays makes the whole process impractical.
If there are automatic tests, they might not reveal that the functionality is broken.
We would then harden the running system, break some functionality, but recognize this much later, probably with some outtakes and user complaints.
Writing good tests is a hard problem, and we did not include the effort of creating such test cases in the effort calculations of our approach.

\subsubsection{Lack of Real Breaking Rule Sets}
\label{subsub:TEST:lack-of-real-breaking-rule-sets}

Our chosen breaking rule sets might not represent breaking rules in real-world systems.
We based our breaking rules sets on results on the literature claim that there are no failures resulting from combining more than $6$ parameters.
However, we do not know what combination of rules causes problems in practice or how their distribution is.
Our assumption here is that the occurring problems might also depend on the size and complexity of the system, i.e., that more complex combinations of breaking rules only occur in more complex systems.
We needed an extensive case study where we collected real-world configuration problems, identified the breaking rules manually, and assessed the complexity.
With this case study's data, we could determine the distribution of the breaking rules and assess how useful our approach is in practice.

\subsubsection{Problematic Proof of Concept Setup}

We assume that one can use our approach with any guide.
However, we show this with our proof of concept implementation only for Windows guides.
Furthermore, we assumed that one can \textit{isolate} the system in an image which might also not be the case for many systems, especially complex systems communicating with each other and the internet.

\subsubsection{Importance of the Maximal Set}

We claim that the algorithm is successful if it returns a maximal set of non-breaking rules for a given profile based on the current breaking rule set.
However, this might not be that important in practice.
If Alice applies all but one rule of the theoretical maximal set, the system's security might not be optimal, but much better than without hardening.
Thus, a more efficient heuristic algorithm returning only a subset of a maximal set might be more economical in practice.
Furthermore, not every rule is equally important as the other.
Thus, we need to incorporate the importance of the rules into our approach.
We need to consider in the future what the optimal solution is and what acceptable solutions are.
The next step will be to adjust the weight of the rules based on the system and its criticality.
The result here could also be that we only need a couple of essential rules on a specific system to reach a good level of security and that the gain of applying the complete guide might not be that high.

\section{Related Work}
\label{sec:TEST:related-work}

Research on configuration management is well-established, but there are constantly new insights into this topic~\cite{velez2022on,dubslaff2022causality,randrianaina2022on,ulhaque2022kgsecconfig}.
Dietrich et al. presented the best overview of the needs of administrators to avoid security misconfigurations~\cite{dietrich2018investigating}.
The work presented in this article depends on previous articles in the context of configuration hardening addressing other essential factors such as the \textit{Poor vendor documentation}~\cite{stockle2022automated}.
Our approach needs automatically implementable guides as presented by \cite{stockle2020automated}.
Furthermore, we use techniques presented in \cite{stockle2022hardening} for the setup and testing of the VMs.

Although originating in the 80s~\cite{mandl1985orthogonal}, the combinatorial testing research is still very active~\cite{wu2020an}.
Kuhn et al. showed that one could use combinatorial testing to test whether the software works with all settings of the software itself, but also whether the software works in every possible configuration of a system~\cite{kuhn2010practical}.
As mentioned before, we use the IPOG~\cite{lei2007ipog} and IPOG-D~\cite{lei2008ipog} algorithm to generate our covering arrays.
The most inspiring article for our work was the approach Yilmaz et al. of to finding faults when testing software using combinatorial test cases~\cite{yilmaz2006covering}.
Furthermore, they suggested further analyzing the test case results using classification tree analysis.
We ported their approach to the domain of security configuration.
However, they were only interested in a combination failing, whereas we are interested in a maximal solution, i.e., applying as many rules as possible.

\section{Conclusion}
\label{sec:TEST:conclusion}

We have shown in this article that one can use combinatorial testing to find combinations of breaking rules and machine-learning-based heuristics to find maximal non-breaking sets $ \mathcal{G} ^ * $.
Administrators can use these sets to harden their systems.
Thus, they will get maximal security from the configuration hardening and keep the system running.

We showed how we could use existing techniques from the software testing domain to solve a problem in configuration hardening.
Since we published our code, we hope administrators can use it to harden their systems.
Furthermore, other researchers can improve our approach or the implementation to devise more efficient ways to harden a system without breaking its functionality.

However, as we have discussed in \autoref{sec:TEST:threats-to-validity}, administrators need automatic tests to apply our approach.
Thus, we advocate for more automatic testing.
Only if administrators have sufficient automatic tests to ensure that all system functions are still working, they will have the courage to implement security measures of any kind.

Until administrators have automatic testing, they can test the covering arrays of the guides in A/B tests:
Then, they apply one tuple to the machines of selected employees.
If employees cannot do their work due to the applied rules, they report this to the administrator.
The administrator marks the tuple as breaking and reverts the rules of the tuple on the employee's machine so they can work again.
After the administrator has tested all tuples, they can use our tool to find the breaking rules based on the breaking tuples.
Ultimately, we will need more testing to have more secure systems in the future.

\bibliographystyle{./IEEEtran}
\bibliography{./main.bib}

\end{document}